\documentclass[aps,prl,amsmath,showpacs,amssymb,floatfix,article]{revtex4}
\usepackage{tabularx}
\usepackage{bm}
\usepackage{subfigure}
\usepackage{euscript}
\usepackage{graphicx}
\usepackage{color}
\usepackage[colorlinks=true,linkcolor=blue]{hyperref}%
\usepackage{amsfonts}
\usepackage{exscale}
\usepackage{amsbsy}

\begin{document}

\title{Topological Superfluid and Majorana Zero Modes in Synthetic Dimension}

\author{Zhongbo Yan$^{1,2}$}
\author{Shaolong Wan$^{2}$}
\author{Zhong Wang$^{1,3}$}
\altaffiliation{  wangzhongemail@tsinghua.edu.cn}
\affiliation{$^{1}$Institute for Advanced
Study, Tsinghua University, Beijing, 100084, China\\
$^{2}$Institute for Theoretical
Physics and Department of Modern Physics University of Science and
Technology of China, Hefei, 230026, P. R. China\\
$^{3}$Collaborative Innovation Center of Quantum Matter,
Beijing 100871, China}


\begin{abstract}
Recently it has been shown that multicomponent spin-orbit-coupled
fermions in one-dimensional optical lattices can be viewed as
spinless fermions moving in two-dimensional synthetic lattices with
synthetic magnetic flux. The quantum Hall edge states in these
systems have been observed in recent experiments. In this paper we
study the effect of an attractive Hubbard interaction. Since the
Hubbard interaction is long-range in the synthetic dimension, it is
able to efficiently induce Cooper pairing between the
counterpropagating chiral edge states. The topological class of the
resultant one-dimensional superfluid is determined by the parity
(even/odd) of the Chern number in the two-dimensional synthetic
lattice. We also show the presence of a chiral symmetry in our model, which implies ${\rm Z}$ classification and the robustness of multiple zero modes when this symmetry is unbroken.
\end{abstract}

\maketitle

Topological superconductors and topological superfluids hosting Majorana
zero modes\cite{b.1,b.2,b.3,b.4,b.5,b.6,b.7} have been among the central themes of
both condensed matter and cold atom physics recently. (``Topological superconductor''
refers to charged particles, while ``topological superfluid'' refers to
neutral particles, otherwise their physics is essentially the same. The
result of our paper is equally applicable to topological superconductors
and topological superfluids.) Apart from being novel phases of matter, they
have potential applications in quantum computation\cite{b.8,b.9}. It is
therefore highly desirable to search for various routes towards realization
of topological superconductivity/superfluidity and Majorana zero modes.  There have
been several proposals to realize them in either condensed matter\cite{b.10,b.11,b.12,
b.13,b.14,b.14,b.15,b.16,b.17,b.18,b.19} or cold atom systems\cite{b.20,b.21,b.22,b.23}.
The latter systems have the advantage of high controllability. Experimental
study on topological superconductors and Majorana zero modes is also extremely active
\cite{b.24,b.25,b.26,b.27,b.28}.

In this paper we study the Cooper pairing between chiral edge modes of quantum Hall
strips as a possible route toward one-dimensional (1D) topological superconductors
and topological superfluids. This is stimulated by the idea of ``synthetic dimension''\cite{b.29,b.30}
emerging from optical lattice with atoms with large
spin\cite{b.31,b.32,b.33,b.34,b.35,b.36,b.37,b.38,b.39,b.40,b.41,b.42}. In this
visualization, internal degrees of freedom (``spin'') form an additional spatial
dimension. This picture is especially convenient when the internal states are coupled
sequentially, as can be readily done by two Raman beams\cite{b.29}. Synthetic magnetic
flux naturally exists inside the two-dimensional  (one physical dimension plus one synthetic
dimension) lattice, and the quantum Hall states (Chern insulator) can be simulated. The
chiral edge modes have been observed in recent experiments\cite{b.43,b.44}. The two
counterpropagating chiral modes are separated in the synthetic dimension, therefore
they are immune to backscattering if the scatters are short-range.

The interaction effect in this system is an important issue\cite{b.29,b.45,b.46,b.47}, both
experimentally and theoretically.  With the motivation of realizing topological superfluidity,
in this paper we study the effect of an attractive Hubbard interaction, which is long-range in the
synthetic dimension, though short-range in the physical dimension. As a consequence, the two
counterpropagating chiral edge modes can be Cooper-paired efficiently.  We find that topological
superfluidity naturally arises. We also study the existence of topological Majorana zero modes. Interestingly, multiple zero modes are stable if a chiral symmetry of our model is unbroken.

\begin{figure}
\includegraphics[height=4.0cm]{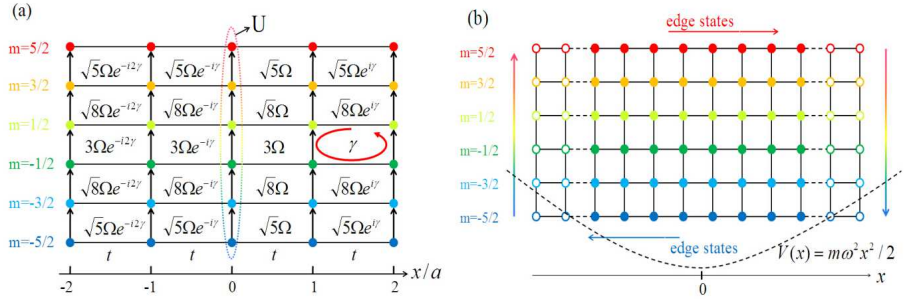}
\caption{ Sketch of the system. The spin states are coupled
sequentially by Raman-induced hopping, which generates a synthetic
dimension, in addition to the physical dimension $x$. }\label{fig1}
\end{figure}

\vspace{2mm}

\noindent
\textbf{Results}

\noindent
\textbf{The model and main picture.}
The system in consideration is illustrated in Fig.\ref{fig1}. The one-dimensional
optical lattice extends in the $x$ direction, and the Raman-induced hopping couples
$M\equiv 2F+1$ spin states at each site $x$ in a sequential manner ($F=5/2$ is shown
in Fig.\ref{fig1}).  Therefore, the system acquires a ``synthetic dimension''\cite{b.29}.
The Hamiltonian is $H=H_0+H_I$, in which the free part is
\begin{eqnarray}
H_{0} &=& \sum_{n,m}(-t
c^\dag_{n+1,m}c_{n,m}+\Omega_{m-1}e^{-i\gamma n}c^\dag_{n,m-1}c_{n,m}
+H.c.) \nonumber \\
&& -\mu \sum_{n,m} c^\dag_{n,m}c_{n,m},
\end{eqnarray}
where $\Omega_m=\Omega  g_{F,m} =\Omega \sqrt{F(F+1)-m(m+1)}$\cite{b.29}, with $\Omega$ depending
on the strength of Raman transitions.  Generally, the value of $\Omega_m$ can be controlled by tuning Raman
beams. Our results will be insensitive to details of $\Omega_m$. The above Hamiltonian is readily realizable in experiment
\cite{b.43,b.44}. The presence of $e^{-i\gamma n}$ indicates that there is a flux $\gamma$
in each plaquette, which is responsible for the emergence of the chiral edge states in this
model(see Fig.(\ref{fig1})). After a gauge transformation $c_{n,m}\rightarrow e^{i\gamma n m} c_{n,m}$,
$H_0$ becomes
\begin{eqnarray}
H_{0} &=& \sum_{n,m}(-t e^{-i\gamma m}
c^\dag_{n+1,m}c_{n,m}+\Omega_{m-1}c^\dag_{n,m-1}c_{n,m}
+H.c.) \nonumber \\
&& -\mu \sum_{n,m} c^\dag_{n,m}c_{n,m},  \label{H}
\end{eqnarray}
in which the hopping gains a spin-dependent phase factor.

In this paper we take the simple yet realistic ${\rm SU}(M)$-invariant Hubbard
interaction:
\begin{eqnarray}
H_{I} =\frac{U}{2}\sum_{n} N_{n}(N_{n}-1)
\end{eqnarray}
where $N_{n}=\sum_m c^\dag_{n,m}
c_{n,m}$ and $U<0$. This interaction is apparently long-range in the
synthetic dimension, thus it is quite capable of pairing the
counterpropagating modes at the opposite edges (near $m=F$ and
$m=-F$, respectively) in the synthetic dimension.

\begin{figure}
\includegraphics[height=4.0cm]{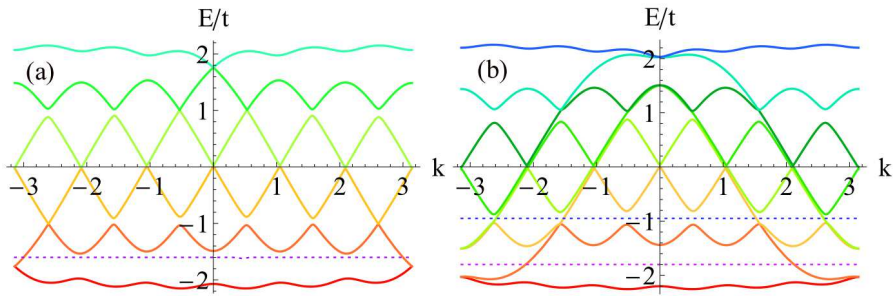}
\caption{ The spectrum of free Hamiltonian $H_0$ for (a) $F=5/2$,
$t=1$, $ \Omega  =0.1$, $\gamma=\pi/3$. The
dotted line is located at $\mu=-1.6$ ; and (b) $F=7/2$, $t=1$,
$\Omega =0.1$, $\gamma=\pi/3$. The purple dotted line is located
at $\mu_1=-1.8$, and the blue dotted line is located at
$\mu_2=-0.95$. }\label{fig2}
\end{figure}

Before proceeding to a quantitative study of $H_I$, we would like to
discuss the physical picture of possible topological superfluidity in
this model.  Since the
hopping lacks translational symmetry along the synthetic dimension,
the bulk Chern number as an integral of Berry
curvature in the two-dimensional Brillouin zone cannot be defined, however, its manifestation as the number of chiral edge
modes is well-defined. Suppose that the ``bulk Chern number'' $C=1$, i.e. there is a single pair of chiral
edge modes in the bulk gap [see Fig.\ref{fig2}, in which the
purple dotted lines intersect with the chiral modes]. If there is a
small Cooper pairing between these two edge modes, the system is a
one-dimensional topological superfluid. This can be inferred using
Kitaev's $Z_2$ topological invariant\cite{b.2}, which
essentially counts the parity (even/odd) of the number of Fermi
points within $[0,\pi/a]$ in the absence of pairing. In the case
$C=2$, i.e. there are two pairs of chiral edge modes, as illustrated
by the curves intersecting with the blue dotted line in
Fig.(\ref{fig2}b), the superfluid resulting from pairing the edge
states is topologically trivial.  Provided that the pairing is small,
the superfluid (or superconductor) is nontrivial(trivial) when the
bulk Chern number is odd(even), namely,
\begin{eqnarray}
(-1)^\nu = (-1)^C,\label{invariant}
\end{eqnarray}
where $\nu=0,1$ (mod $2$) is the ${\rm Z}_2$
topological number of 1D superconductor or
superfluid\cite{b.2}. In the rest part of this paper we shall
present a quantitative study of the picture outlined above.

\begin{figure}
\includegraphics[height=4.0cm]{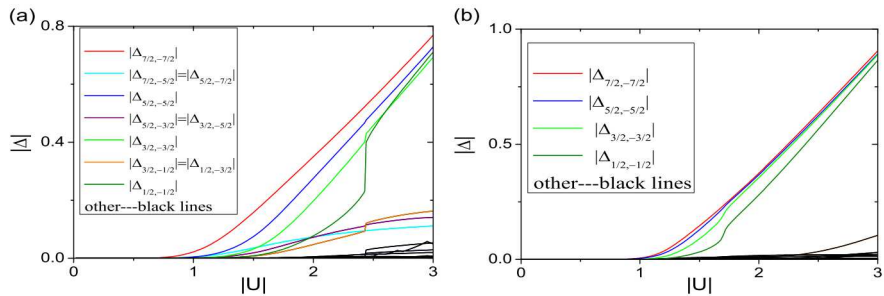}
\caption{   The self-consistent solutions of Cooper pairing
$\Delta_{mm'}$ as functions of $U$ for (a) $\mu_1=-1.8$; (b) $\mu_2=-0.95$.  The parameters are $t=1$,
$\Omega =0.1$, and $\gamma=\pi/3$, which are the same as used in
Fig.(\ref{fig2}b).
  }\label{fig3}
\end{figure}

\vspace{2mm}

\noindent
\textbf{Cooper pairing in self-consistent mean-field.} At the mean-field level the
Hubbard interaction can be decomposed as
\begin{eqnarray}
H_{I} = \frac{1}{2}
\sum_{n,mm'} (\Delta_{n,mm'}c^\dag_{n,m}c^\dag_{n,m'} +
\Delta^*_{n,mm'}c_{n,m'}c_{n,m}) +
\sum_{n,mm'}\frac{|\Delta_{n,mm'}|^2}{2|U|},
\end{eqnarray} where we have
defined $\Delta_{n,mm'}= U<c_{nm'}c_{nm}>$, which satisfies
$\Delta_{n,mm'}= -\Delta_{n,m'm}$ because of Fermi statistics. The
mean field BdG Hamiltonian becomes
\begin{eqnarray}
H_{{\rm MF}} = H_0+
\frac{1}{2} \sum_{n,mm'} (\Delta_{n,mm'}c^\dag_{n,m}c^\dag_{n,m'} +
\Delta^*_{n,mm'}c_{n,m'}c_{n,m}).  \label{mean-field}
\end{eqnarray}
Since the
basic physics is the pairing of chiral edge modes with opposite
momenta, it is natural to consider Cooper pairing with zero total
momentum, namely that $\Delta_{n,mm'}\equiv \Delta_{mm'}$ is
independent of $n$.  We determine these $\Delta_{mm'}$
self-consistently. Whenever there are several sets of self-consistent
solutions of $\{\Delta_{mm'}\}$, we compare their mean-field energies
and pick up the ground state. This calculation can be carried out for
all possible values of $F$. Hereafter we take $F=7/2$ as an example.

The pairings as functions of Hubbard $U$ are shown in
Fig.(\ref{fig3}) for two values of chemical potential
$\mu$. In Fig.(\ref{fig3}a) we take $\mu_1=-1.8$, which
corresponds to bulk Chern number $C=1$  [see the purple dotted line
in Fig.(\ref{fig2}b)]. For small $|U|$, as $|U|$ increases,
$\Delta_{mm'}$ grows exponentially. For $U$ not too large,
$\Delta_{7/2,-7/2}$ dominates other pairings, which is consistent
with the picture of pairing between chiral edge modes. At a critical
$|U|$ slightly below $2.5$, the system undergoes a first-order
transition to a phase in which $\Delta_{1/2,-1/2}$, which should be
regarded as a ``bulk pairing'', becomes comparable to the edge
pairing. Thus the edge-pairing picture becomes inaccurate at large
$|U|$. In Fig.(\ref{fig3}b) a different chemical potential
$\mu_2=-0.95$ is taken, and the edge-state pairings again dominate at
small $|U|$.

\begin{figure}
\includegraphics[height=4.0cm]{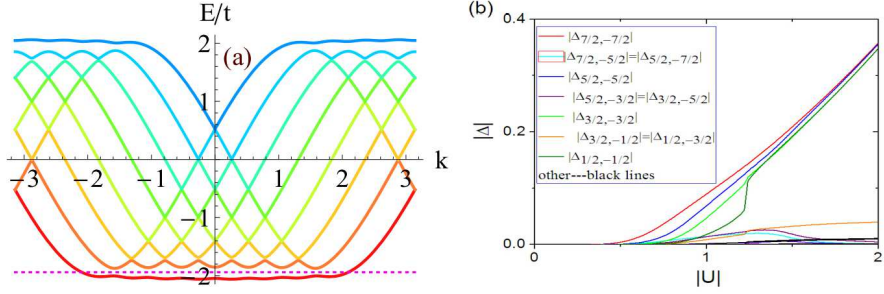}
\caption{ (a) The spectrum without interaction for $t=1$,
$\Omega=0.025$, and $\gamma=\pi/6$. The dashed line marks
$\mu=-1.94$. (b) The Cooper pairings as functions of $U$.
}\label{fig4}
\end{figure}

To illustrate the robustness of the present picture, we study a set of quite different
parameters, shown in Fig.(\ref{fig4}). The behavior of pairings is qualitatively
the same as that for the previous parameters.

A few remarks are in order. First, although we have taken the strength of hopping along
the synthetic dimension as $\Omega g_{F,m}$, we have also checked that the result is
qualitatively the same when it is $m$-independent. Second, when the 2D bulk is metallic,
topological superfluidity can still emerge, though there is no clear criteria using Chern
number. We shall not focus on details about this.

\vspace{2mm}

\noindent
\textbf{Majorana zero modes.} The hallmark of 1D topological superconductor or superfluid
is the emergence of topological zero modes localized near the two ends of an open chain.

\begin{figure}
\includegraphics[height=4.0cm]{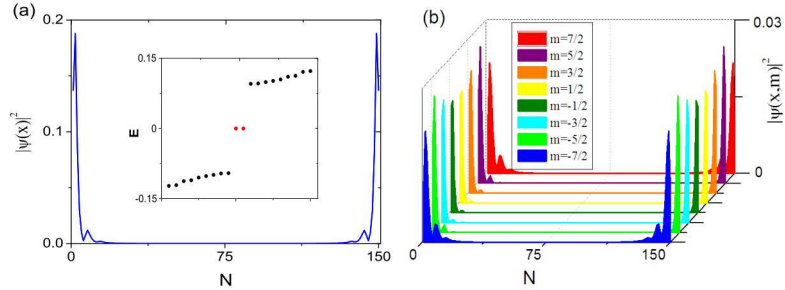}
\caption{ (a) The wavefunction of a zero mode in an open chain with
length $N=150$. The parameters are $t=1,\Omega=0.1,\gamma=\pi/3$ and
$\mu=-1.8$. The Cooper pairings are taken as $\Delta_{7/2,-7/2}=0.09,
\Delta_{7/2,-5/2}=-0.019, \Delta_{5/2,-5/2}=0.03,
\Delta_{5/2,-3/2}=-0.007, \Delta_{3/2,-3/2}=0.01,
\Delta_{3/2,-1/2}=-0.002, \Delta_{1/2,-1/2}=0.002$, which are the
mean-field pairing obtained at $|U|=1.3$ ( see also
Fig.(\ref{fig3}a) ). Other $\Delta_{mm'}$'s are much
smaller and thus neglected. The two zero modes have the same profile
of $|\psi(x)|^2 \equiv \sum_m|\psi(x,m)|^2$, thus only one is shown
here. The inset shows a few energies near $E=0$. (b) The zero mode
solution with $m$ resolution.  }\label{fig5}
\end{figure}

\begin{figure}
\includegraphics[width=16cm, height=4.0cm]{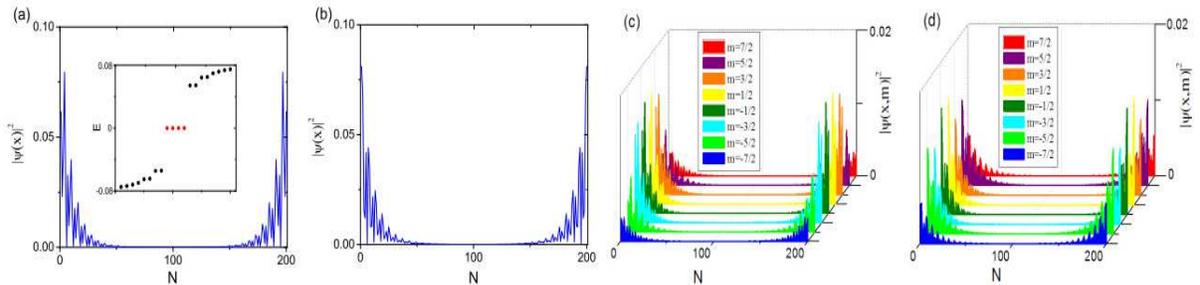}
\caption{ The wavefunction of zero modes in an open chain with length
$N=200$. The parameters are $t=1,\Omega=0.1,\gamma=\pi/3$ and
$\mu=-0.95$, which is the same as Fig.\ref{fig5} except for the
chemical potential.  The pairings are $\Delta_{7/2,-7/2}=0.067,
\Delta_{7/2,-5/2}=-0.005, \Delta_{5/2,-5/2}=0.052,
\Delta_{3/2,-3/2}=0.028, \Delta_{1/2,-1/2}=0.009$ (the mean-field
pairing at $U=-1.3$), and other pairing terms are much smaller and
thus neglected. The spatial profile of $|\psi(x)|^2$ of two of the
four zero modes are shown in (a) and (b), while the profile with $m$
resolution is shown in (c) and (d). The other two zero modes with the
same profiles are not shown repeatedly. The inset of (a) shows
several energies close to $E=0$. }\label{fig6}
\end{figure}

We have solved the BdG mean-field Hamiltonian Eq.(\ref{mean-field})
for the wavefunction $\psi(x,m)=[u(x,m),v(x,m)]^T$, in which $u(x,m)$
and $v(x,m)$ denotes the particle and hole component respectively.
Below we present our solutions in a chain with sharp boundary, for
the parameters $t=1$, $\Omega =0.1$, $\gamma=\pi/3$, at both
$\mu_1=-1.8$ and $\mu_2=-0.95$. The Cooper pairings are taken to be
the mean-field values at $U=-1.3$, which we obtained in the previous
section. The case $\mu_1=-1.8$ is shown in Fig.\ref{fig5}. There
is one zero mode localized at each end of the open chain, and a tiny
finite-size coupling mixes them slightly, though the energy splitting
due to finite-size effect is too small to be discernable. The
existence of a single Majorana zero modes at each end of an open
chain is consistent with Eq.(\ref{invariant}), the bulk Chern number
being $C=1$ (odd number) at $\mu_1=-1.8$. The superfluid is
topologically nontrivial in this case.

As a comparison, we also present the zero mode solutions at
$\mu_2=-0.95$, for which the free Hamiltonian have two pairs of
chiral edge modes ($C=2$), indicating that the superfluid phase
at small $|U|$ should be $Z_2$ topologically trivial (see
Eq.(\ref{invariant})). In the numerical calculation with open
boundary condition, we find two Majorana zero modes at each end
(see Fig.\ref{fig6}),
which means that the superfluid is ${\rm Z}_2$ trivial. Therefore
we see again that $(-1)^C$ determines the $Z_2$
topological classification of the 1D superfluid in synthetic
dimension.

One may wonder why there is no hybridization between the two zero modes, which may open a gap for them. We shall explain the reason as follows.
In fact, the BdG Hamiltonian has a time-reversal symmetry and a particle-hole symmetry, which can be combined into a chiral symmetry\cite{b.49}. If the Cooper pairing $\Delta_{mm'}$ are real, then we can check that the Hamiltonian satisfies
\begin{eqnarray}
CH_{{\rm MF}}(k)C^{-1}= -H_{{\rm MF}}(k)
\end{eqnarray}
 in which
 \begin{eqnarray}
C = \left(
                \begin{array}{ccccc}
                    &  &  &  & -i \\
                   &  &  & -i &  \\
                   & &  \cdots &  &  \\
                   & i &  &   &  \\
                  i &  & & &   \\
                \end{array}
 \right).
\end{eqnarray}
This matrix is written in the BdG basis of $(c_{k,F},\cdots,c_{k,-F},c^\dag_{-k,F},\cdots,c^\dag_{-k,-F})$.
Due to these symmetries, the BdG Hamiltonian can be classified as
BDI\cite{b.48,b.49}, whose classification in 1D is ${\rm Z}$. The $Z_2$ topologically trivial phase is nontrivial according to the $Z$ classification of BDI class, which is the reason why zero modes appear at the edge of $Z_2$ trivial states.  We have
checked that, if we break the symmetries, e.g. by giving a phase factor
to $\Delta_{7/2,-7/2}$ (with other $\Delta_{mm'}$s unchanged), then
these zero modes will be shifted to nonzero energies.

According to Fidkowski and Kitaev's work\cite{b.52}, in the presence of interaction, the classification of BDI-class topological superconductors in 1D is Z$_8$ instead of Z. Because of the flexible tunability to topological superconductors with large topological number (say $8$) in our system, the Z$_8$ classification can be tested experimentally. If we tune the bulk Chern number of our system to $8$, the eight nominally-zero-modes will be shifted away from zero energy due to the (beyond-mean-field) interaction effects of the Hubbard term. If observed, this will be an experimental test of Fidkowski and Kitaev's Z$_8$ classification.

\begin{figure}
\includegraphics[height=4.0cm]{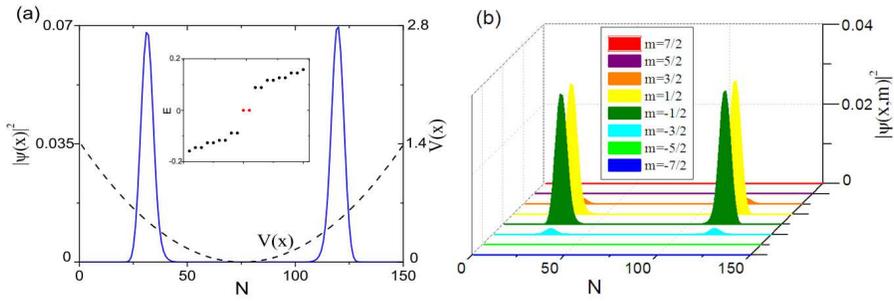}
\caption{  Majorana zero modes in the harmonic trap.   The parameters
are $t=1$, $\Omega =0.1$, and $\gamma=\pi/3$, which are the same as
used in Fig.(\ref{fig2}b). The harmonic trap is $V(x) = 0.00025
(x/a)^2$, where $x=0$ is the center of the chain with size $N=150$.
The two zero modes have the same profile of $|\psi(x,m)|^2$, thus we
only show one. The inset of (a) shows several energies close to
$E=0$.
  }\label{fig7}
\end{figure}

To make a closer connection to experiment, we also study the
existence of Majorana zero modes in a system with soft boundary
created by a harmonic trap $V(x)$.  In the presence of $V(x)$,  the
chemical potential becomes $\mu(x) = \mu_0- V(x)$. We take
$\mu_0=-1.7$ and $V(x) = 0.00025 (x/a)^2$, such that the center of
the system is topologically nontrivial. Since $\mu(x)$ is not
constant, $\Delta_{mm'}$ should also be $x$-dependent. To incorporate
this effect, we numerically calculate the functions
$\Delta_{mm'}(\mu)$ at $U=-1.3$, which is then used to produce the
mean-field BdG Hamiltonian in harmonic trap. In the solution to this
BdG Hamiltonian, the zero modes can be clearly seen, as shown in
Fig.(\ref{fig7}), though the quantitative details are different
from the case of hard boundary.

\vspace{2mm}

\noindent
\textbf{Conclusions and Discussions.}

\noindent
We have studied the pairing between counterpropagating chiral edge modes in
the quantum Hall strip in synthetic dimension. This picture has several merits.
Creation of magnetic flux in the synthetic dimension by Raman beans is easier
than in physical dimensions. The spatial separation of left and right moving
chiral edge states in the synthetic dimension effectively prevents the
backscattering between them, which implies their robustness. Meanwhile, the
Hubbard interaction is not suppressed by this spatial separation: It is
infinite-range in the synthetic dimension, therefore, it can pair
the two edge modes quite efficiently. As we have shown, the resultant states are
topological superfluid carrying Majorana zero modes. If the chiral symmetry of our model is unbroken, the classification is ${\rm Z}$ and multiple zero modes are stable; on the other hand, if this symmetry is broken, the classification is ${\rm Z}_2$.

Finally, we remark that quantum fluctuations of the phase factor of pairing in 1D
is generally strong.  One can put the 1D system in proximity to a 3D supefluid to
suppress these fluctuations\cite{b.20}. Moreover, it has been shown\cite{b.50,b.51}
that long-range superconducting order is not a necessary condition for the existence
of Majorana zero modes. The zero modes persist even when the long-range superconducting
order is replaced by algebraic order (i.e. the correlations of pairings decay by power-law).
In our system this conclusion applies.

\vspace{2mm}

\noindent
\textbf{Methods}

\noindent
\textbf{Mean-field calculations.}
The mean-field calculation is carried out by the standard procedure of decomposing the Hubbard interaction as fermion bilinear terms,   leading to Eq.(\ref{mean-field}). The Cooper pairing is calculated from Eq.(\ref{mean-field}) in a self-consistent manner. All self-consistent solutions for the Cooper pairing are obtained. In the case that there are more than one self-consistent solutions, the one with lowest mean-field energy is selected.

\vspace{2mm}

\noindent
\textbf{Acknowledgments}

\noindent
We are especially grateful to Hui Zhai for stimulating discussions.
This work is supported by NSFC under Grant No. 11304175, No.11275180,
and Tsinghua University Initiative Scientific Research Program.

%
%
%
%
%


\begin{thebibliography}{99}

\bibitem{b.1} Read, N. \& Green, D. Paired states of fermions in two dimensions with breaking of parity and time-reversal symmetries and the fractional quantum Hall effect. \textit{Phys. Rev. B} {\bf61}, 10267 (2000).

\bibitem{b.2} Kitaev, A. Y. Unpaired Majorana fermions in quantum wires. \textit{Physics-Uspekhi} {\bf44}, 131 (2001).

\bibitem{b.3} Qi, X. L. \& Zhang, S. C. Topological insulators and superconductors. \textit{Rev. Mod. Phys.} {\bf83}, 1057 (2011).

\bibitem{b.4} Alicea, J. New directions in the pursuit of Majorana fermions in solid state systems. \textit{Rep. Prog. Phys.}
{\bf75}, 076501 (2012).

\bibitem{b.5} Beenakker, C. W. J. Search for Majorana fermions in superconductors. \textit{Annu. Rev. Con. Mat. Phys.} {\bf4},
113 (2013).

\bibitem{b.6} Wilczek, F. Majorana returns. \textit{Nat. Phys.} {\bf5}, 614 (2009).

\bibitem{b.7} Elliott, S. R. \& Franz, M. Colloquium: Majorana Fermions in nuclear, particle and solid-state physics. \textit{Rev. Mod. Phys.} {\bf87}, 137 (2015).

\bibitem{b.8} Kitaev, A. Y.  Fault-tolerant quantum computation by anyons. \textit{Ann. Phys.} {\bf303}, 2 (2003).

\bibitem{b.9} Nayak, C., Simon, S. H., Stern, A., Freedman, M., \& Das Sarma, S. Non-Abelian Anyons and Topological Quantum Computation. \textit{Rev. Mod. Phys.} {\bf80}, 1083 (2008).

\bibitem{b.10} Fu, L. \& Kane, C. L. Superconducting Proximity Effect and Majorana Fermions at the Surface of a Topological Insulator. \textit{Phys. Rev. Lett.} {\bf100}, 096407 (2008).

\bibitem{b.11} Lutchyn, R. M., Sau, J. D. \& Das Sarma, S. Majorana Fermions and a Topological Phase Transition in Semiconductor-Superconductor Heterostructures. \textit{Phys. Rev. Lett.} {\bf105}, 077001 (2010).

\bibitem{b.12} Oreg, Y.,  Refael, G.  \&  von Oppen F. Helical Liquids and Majorana Bound States in Quantum Wires. \textit{Phys. Rev. Lett.} {\bf105}, 177002 (2010).

\bibitem{b.13} Sau, J. D., Lutchyn, R. M., Tewari, S. \& Das Sarma, S. Generic New Platform for Topological Quantum Computation Using Semiconductor Heterostructures. \textit{Phys. Rev. Lett.} {\bf104}, 040502 (2010).

\bibitem{b.14} Tewari, S., Das Sarma, S., Nayak, C., Zhang, C. \& Zoller, P. Quantum Computation using Vortices and Majorana Zero Modes of a $p_{x}+ip_{y}$ Superfluid of Fermionic Cold Atoms. \textit{Phys. Rev. Lett.} {\bf98}, 010506 (2007).

\bibitem{b.15} Alicea, J. Majorana fermions in a tunable semiconductor device. \textit{Phys. Rev. B} {\bf81}, 125318 (2010).

\bibitem{b.16} Stanescu, T. D., Lutchyn, R. M. \& Das Sarma, S. Majorana fermions in semiconductor nanowires. \textit{Phys. Rev. B} {\bf84}, 144522 (2011).

\bibitem{b.17} Alicea, J., Oreg, Y., Refael, G., von Oppen, F. \& Fisher, M. P. A. Non-Abelian statistics and topological quantum information processing in 1D wire networks. \textit{Nat. Phys.} {\bf7}, 412 (2011).

\bibitem{b.18} Qi, X. L., Hughes, T. L. \& Zhang, S. C. Chiral topological superconductor from the quantum Hall state. \textit{Phys. Rev. B} {\bf82}, 184516 (2010).

\bibitem{b.19} Potter, A.  C.  \&  Lee, P.  A. Multichannel Generalization of Kitaev¡¯s Majorana End States and a Practical Route to Realize Them in Thin Films. \textit{Phys. Rev. Lett.} {\bf105}, 227003 (2010).


\bibitem{b.20} Jiang, L. et al. Majorana Fermions in Equilibrium and in Driven Cold-Atom Quantum Wires. \textit{Phys. Rev. Lett.} {\bf106}, 220402 (2011).

\bibitem{b.21} Zhang, C., Tewari, S., Lutchyn, R. M. \& Das Sarma, S. $p_{x}+ip_{y}$ Superfluid from s-Wave Interactions of Fermionic Cold Atoms. \textit{Phys. Rev. Lett.} {\bf101}, 160401 (2008).

\bibitem{b.22} Sato, M., Takahashi, Y.  \& Fujimoto, S. Non-Abelian Topological Order in $s$-Wave Superfluids of Ultracold Fermionic Atoms. \textit{Phys. Rev. Lett.} {\bf103}, 020401 (2009).

\bibitem{b.23} Diehl, S., Rico, E., Baranov, M. A. \& Zoller, P. Topology by dissipation in atomic quantum wires. \textit{Nat. Phys.} {\bf7}, 971 (2011).

\bibitem{b.24} Mourik, V. et al.  Signatures of Majorana Fermions in Hybrid Superconductor-Semiconductor Nanowire Devices. \textit{Science} {\bf336}, 1003 (2012).

\bibitem{b.25} Nadj-Perge, S. et al.  Observation of Majorana fermions in ferromagnetic atomic chains on a superconductor. \textit{Science} {\bf346}, 602 (2014).

\bibitem{b.26} Das, A. et al.  H. Zero-bias peaks and splitting in an Al¨CInAs nanowire topological superconductor as a signature of Majorana fermions. \textit{Nat. Phys.} {\bf8}, 887 (2012).

\bibitem{b.27} Wang, M. X. et al. The Coexistence of Superconductivity and Topological Order in the Bi2Se3 Thin Films. \textit{Science} {\bf336}, 52 (2012).

\bibitem{b.28} Finck, A. D. K. et al. Anomalous Modulation of a Zero-Bias Peak in a Hybrid Nanowire-Superconductor Device. \textit{Phys. Rev. Lett.} {\bf110}, 126406 (2013).

\bibitem{b.29} Celi, A. et al. Synthetic gauge fields in synthetic dimensions. \textit{Phys. Rev. Lett.} {\bf112}, 043001 (2014).

\bibitem{b.30} Boada, O., Celi, A., Latorre, J. I. \& Lewenstein, M. Quantum Simulation of an Extra Dimension. \textit{Phys. Rev. Lett.} {\bf108}, 133001 (2012).

\bibitem{b.31} Gorshkov, A. et al. Two-orbital $SU(N)$ magnetism with ultracold alkaline-earth atoms. \textit{Nat. Phys.}, {\bf 6}, 289 (2010).

\bibitem{b.32} Taie, S. et al. Realization of a $SU(2)\times SU(6)$ System of Fermions in a Cold Atomic Gas. \textit{Phys. Rev. Lett.} {\bf105}, 190401 (2010).

\bibitem{b.33} Hara, H., Takasu, Y., Yamaoka, Y., Doyle, J. M. \& Takahashi, Y. Quantum Degenerate Mixtures of Alkali and Alkaline-Earth-Like Atoms.
  \text{Phys. Rev. Lett.} {\bf106}, 205304 (2011).

\bibitem{b.34} DeSalvo, B. J., Yan, M., Mickelson, P. G., Martinez de Escobar, Y.  N. \& Killian, T.  C. Degenerate Fermi Gas of $^87$Sr. \textit{Phys. Rev. Lett.} {\bf105}, 030402 (2010).

\bibitem{b.35} Scazza, F. et al. Observation of two-orbital spin-exchange interactions with ultracold $SU(N)$-symmetric fermions
  \textit{Nat. Phys.} {\bf10}, 779 (2014).

\bibitem{b.36} Ho, T. L. \& Yip, S. Pairing of Fermions with Arbitrary Spin. \textit{Phys. Rev. Lett.} {\bf 82}, 247 (1999).

\bibitem{b.37} Wu, C. J., Hu, J. P. \&  Zhang, S. C. Exact $SO(5)$ Symmetry in the Spin-$3/2$ Fermionic System.
  \textit{Phys. Rev. Lett.} {\bf91}, 186402 (2003).

\bibitem{b.38} Wu, C. J. Hidden symmetry and quantum phases in spin-$3/2$ cold atomic systems. \textit{Mod. Phys. Lett. B } {\bf 20}, 1707 (2006).

\bibitem{b.39} Lecheminant, P., Boulat, E. \& Azaria P. Confinement and Superfluidity in One-Dimensional Degenerate Fermionic Cold Atoms. \textit{Phys. Rev. Lett.} {\bf95}, 240402 (2005).

\bibitem{b.40} Hermele, M., Gurarie, V.  \& Rey, A.  M. Mott Insulators of Ultracold Fermionic Alkaline Earth Atoms: Underconstrained Magnetism and Chiral Spin Liquid. \textit{Phys. Rev. Lett.} {\bf103}, 135301 (2009).

\bibitem{b.41} Zhang, X. et al. Spectroscopic observation of $SU(N)$-symmetric interactions in Sr orbital magnetism. \textit{Science} {\bf345}, 1467 (2014).

\bibitem{b.42} Cazalilla, M. A. \& Rey, A. M. Ultracold Fermi gases with emergent SU(N) symmetry.
 \textit{Rep. Prog. Phys.} {\bf77}, 124401 (2014).

\bibitem{b.43} Stuhl, B. K., Lu, H. I., Aycock, L. M., Genkina, D. \& Spielman, I. B.  Visualizing edge states with an atomic Bose gas in the quantum Hall regime. \textit{arXiv: 1502.02496}.

\bibitem{b.44} Mancini, M. et al. Observation of chiral edge states with neutral fermions in synthetic Hall ribbons. \textit{arXiv: 1502.02495}.

\bibitem{b.45} Barbarino, S., Taddia, L., Rossini, D., Mazza, L. \& Fazio, R.  Magnetic crystals and helical liquids in alkaline-earth fermionic gases. \textit{arXiv: 1504.00164}.

\bibitem{b.46} Ghosh, S. K., Yadav, U. K. \& Shenoy, V. B. Baryon squishing in synthetic dimensions by effective SU(M) gauge fields. \textit{arXiv: 1503.02301}.

\bibitem{b.47} Zeng, T. S., Wang, C. \& Zhai, H.  Charge Pumping of Interacting Fermion Atoms in the Synthetic Dimension. \textit{arXiv: 1504.02263}.

\bibitem{b.48} Schnyder, A. P., Ryu, S., Furusaki, A., \& Ludwig, A. W. W. Classification of topological insulators and superconductors in three spatial dimensions, \textit{Phys. Rev. B} {\bf78}, 195125 (2008).

\bibitem{b.49} Ryu, S., Schnyder, A. P., Furusaki, A., \& A. W. W. Ludwig, Topological insulators and superconductors: tenfold way and dimensional hierarchy, \textit{New. J. Phys.} {\bf12}, 065010 (2010).

\bibitem{b.50} Fidkowski, L., Lutchyn, R. M., Nayak, C. \& Fisher, M. P. A. Majorana zero modes in one-dimensional quantum wires without long-ranged superconducting order. \textit{Phys. Rev. B} {\bf84}, 195436 (2011).

\bibitem{b.51} Ruhman, J., Berg, E.   \&  Altman, E. Topological States in a One-Dimensional Fermi Gas with Attractive Interaction. \textit{Phys. Rev. Lett.} {\bf114}, 100401 (2015).

\bibitem{b.52} Fidkowski, L., \& Kitaev, A.  Effects of interactions on the topological classification of free fermion systems. \textit{Phys. Rev. B}, 81, 134509  (2010).

\end{thebibliography}
\end{document}